\documentclass[onecolumn,usenatbib,usegraphicx]{mn2e}
\def\mnras{MNRAS}
\def\apj{ApJ}
\def\apjl{ApJL}
\def\LL{LL}

\def\k{{\bf k}}
\def\n{{\bf \hat{n}}}
\def\xib{\bar{\xi}}
\def\ss{(s_{\perp},\st)}
\def\s{{\bf s}}
\def\r{{\bf r}}

\def\spp{(s_{\perp},\st+\st^{'})}
\def\st{s_{\parallel}}
\def\spt{s^{'}_{\parallel}}
\def\v{{\bf v}}

\def\vp{V_P}
\def\sig{\sigma_P^2}
\def\sigm{\sigma_P}
\usepackage{epsfig}

\begin{document}
\title[Modeling non-linear effects in  redshift space]
{ Modeling non-linear effects in the redshift space two-point
  correlation function and its implications for the pairwise velocity
  dispersion}  
\author[B. Pandey and S. Bharadwaj]{ Biswajit Pandey\thanks{Email:
 pandey@cts.iitkgp.ernet.in} and Somnath Bharadwaj\thanks{Email: 
    somnathb@iitkgp.ac.in} 
\\ Department of Physics and Meteorology\\ and
\\ Centre for Theoretical Studies \\ IIT Kharagpur \\ Pin: 721 302 ,
India }
\maketitle
\begin{abstract}
The anisotropies in the galaxy two-point correlation function measured
from  redshift surveys exhibits deviations from the predictions of the
linear theory of redshift space distortion on scales as large as $20\,
h^{-1} \, {\rm Mpc}$  where we expect linear theory to hold in real
space.  Any attempt at analyzing the anisotropies in the redshift
correlation function and determining the linear  distortion
parameter $\beta$ requires these deviations  to  be correctly modeled
and taken into account. These deviations are usually
attributed to galaxy random motions and these are incorporated in the 
analysis through a phenomenological model where the linear redshift
correlation is convolved with the random pairwise  velocity
distribution  function along the line of sight. We show that a
substantial part of the deviations  arise from non-linear
effects in the mapping from real to redshift space caused  by the 
coherent flows. Models which incorporate this effect provide an
equally good  fit to N-body results as compared to the
phenomenological model which has only the effect of random motions.
We find that the pairwise velocity
dispersion  predicted by all the models that we have considered are
in excess of the values determined directly from the N-body
simulations. This indicates a shortcoming in our understanding of the
statistical properties of peculiar velocities and their relation to
redshift distortion. 
\end{abstract}


\section{Introduction}

Galaxy redshifts are not perfectly described by pure Hubble's
law. Density fluctuations  induce peculiar velocities relative to the
general Hubble expansion. The peculiar velocities perturb galaxy
redshifts which in turn affects  their inferred distances and this
leads to a systematic distortion in 
the clustering pattern of galaxies in redshift space.
The peculiar velocities cause the two-point correlation function in
redshift space $ \xi^s( \s)$   to be 
anisotropic {\it i.e.} it depends separately on  the
component of the pair separation  $\s$  parallel $ (s_{\|}) $ and
perpendicular $(s_\bot )$ to the observer's line of sight $\n$. There
are 
two characteristic effects of 
peculiar velocities . On large scales structures are compressed along
the line of sight due to coherent flows into over dense regions and
out of under dense regions, thereby amplifying $ \xi^s ( s_{\|}, s_\bot
)$. On small scales $  \xi^s ( s_{\|}, s_\bot)$ is suppressed due to
the structures being elongated along the line of sight by random
motions in virialized clusters.

\citet{kais} first quantified the correlation anisotropy that
 results from large-scale peculiar flows in terms of the power
 spectrum of galaxy clustering.  Using linear theory and the plane
 parallel approximation he showed that the power
spectrum in redshift space $P_s(\k)$ and it's real space counterpart
$P_r(k) $ are related as 
\begin{equation}
P_s(\k) = ( 1+ \beta \mu ^2_k )^2 \ P_r (k) 
\end{equation}
where $ \mu_k $ is the cosine of the angle between $\k$ and the line of
sight $\n$,  and $ \beta \simeq \Omega_m^{0.6}/b$ is the linear
distortion parameter. Here $\Omega_m$ is the cosmic mass density
parameter and b is the linear bias parameter which 
differs from  unity if the galaxies represent a biased sample of the
underlying dark matter distribution. It may be noted that the factor
$\Omega_m^{0.6}$ relates peculiar velocities to density  density
fluctuations \citep{peeb}. This is slightly modified in the
presence of a cosmological constant \citep{lahav1} and it is more
accurate to use $\beta=f(\Omega_m)/b$ where 
$
f(\Omega_m)=\Omega_m^{0.6}+\frac{1}{70}[1-\frac{1}{2}\Omega_m(1+\Omega_m)]
\,. $
The important point is that the
anisotropies observed in $P_s(\k)$ can be used to determine the value
of $\beta$, and thereby place interesting constraints on the density
parameter $\Omega_m$  and the bias $b$. This has been the single most
important motivation for a substantial amount of the research which
has been carried out in trying to understand and quantify the nature
of redshift space distortions. 

\citet{ham} translated Kaiser's linear formula from Fourier to real
space. 
He showed that it is most convenient to parameterize the anisotropy of 
$ \xi ^s ( s_\parallel , s_\perp)$ in terms of spherical harmonics as 
\begin{equation}
\xi^s (s_\parallel , s_\perp) = \sum_{l=0}^{\infty} \xi_l (s) P_l
(\mu )
\end{equation}
where $s=\sqrt{s^2_\parallel+s^2_\perp}$, $\mu=s_\parallel/s$,
$P_l(\mu)$ are the Legendre polynomials and $\xi_l(s)$ are the
different angular moments of the redshift space two-point correlation
function. Only the first three even angular moments, namely the
monopole $\xi_0(s)$, the quadrupole $\xi_2(s)$ and the hexadecapole
$\xi_4(s)$ are non-zero and these can be expressed in terms of the 
real space galaxy two-point correlation $\xi(r)$ and its moments
which are defined as  

\begin{equation}
\xib_n(s)=\frac{n+1}{s^{n+1}} \int_0^s \xi(y) y^n dy\,.
\label{eq:a2}
\end{equation}
through 
\begin{equation}
\xi_0(s)=(1+\frac{2}{3} \beta  + \frac{1}{5} \beta^2) \,  \xi(s)
\label{eq:a3}
\end{equation}
\begin{equation}
\xi_2(s)= (\frac{4}{3} \beta + \frac{4}{7} \beta^2) \, 
\left[\xi(s)-\xib_2(s) \right]
\label{eq:a4}
\end{equation}
\begin{equation}
\xi_4(s)=\frac{8}{35} \beta^2 \, [\xi(s) + \frac{5}{2} \xib_2(s)
-\frac{7}{2} \xib_4(s) ]
\label{eq:a5}
\end{equation}
The linear analysis predicts a negative quadrupole ($i.e. \, \xi_2(s)
<0$) arising from the squashing of large scale structures along the
line of sight. 

Hamilton proposed that the observed redshift space correlation
function be decomposed into spherical harmonics, and the  ratio 

\begin{equation}
Q(s)=\frac{\xi_2(s)}{\frac{3}{s^3}\int_0^s \xi_0(s^{'}) d s^{'}-
  \xi_0(s)} = \left[ \frac{\frac{4}{3} \beta + \frac{4}{7} \beta^2} 
{1+\frac{2}{3} \beta  + \frac{1}{5} \beta^2}\right]
\label{eq:a6}
\end{equation}
which is expected to have  a constant value  
%in the linear regime, 
(shown in $[...]$ in eq. \ref{eq:a6}) 
be used to determine the value of $\beta$. Alternatively, if the real
space correlation function has a power law behaviour $\xi(r)\propto 
r^{-\gamma}$, the ratio $\xi_2(s)/\xi_0(s)$ is also expected to be a
constant, and this can be used to determine the value of $\beta$. 

%has sometimes  been used to determine the value of
%$\beta$ (eg. \citealt{pcok}). 

Investigations using N-body simulations to study the redshift space
two-point correlation (eg. \citealt{suto},  \citealt{fish94},
\citealt{brain},  \citealt{brom}) find deviations from 
the linear predictions out to scales as large as $20 \, h^{-1} \, {\rm
  Mpc}$ and even larger where linear theory is known to be valid in  
real space. Such deviations are also seen in the redshift space
two-point correlations determined from different redshift surveys 
(eg. \citealt{landy} (LCRS), \citealt{pcok} (2dFGRS), \citealt{haw}
(2dFGRS)).  In 
addition to the squashing predicted by the linear
analysis,  the two-point correlation  function exhibits an elongation
along the line of sight at scales as large as  $ 20 h^{-1} $ Mpc. This
causes the quadrupole moment to remain positive even at scales 
where one would expect linear theory to hold in real space. 
The values of $Q(s)$ which are expected to be constant
(eq. \ref{eq:a6}) do not flatten out to scales as large as $20 \,
h^{-1} \, {\rm Mpc}$ in N-body simulations, nor is the flattening
observed at these scales in the redshift surveys. All this indicates
that there are 
non-linear effects which are important in the mapping from real space
to redshift space at length-scales where linear theory is known to be
valid in real space. 

The most popular approach is to attribute the deviations from the
linear predictions to the effects of the random  
peculiar velocities of galaxies located in virialized clusters and
other highly non-linear regions. This effect is incorporated through a
phenomenological model (eg. \citealt{dav83}, \citealt{fish94},
\citealt{pcokanddodds},  \citealt{hevtayl}, \citealt{marj},
\citealt{bali}, \citealt{tadefsta},
\citealt{brom}, \citealt{rat}) which assumes that
at large scales the deviations from linear theory can be incorporated
by convolving the  
linear redshift space correlation function $ \xi ^s _L $  with the line
of sight component of the random, isotropic pairwise velocity
distribution function  ${\it f}(\upsilon)$. The resulting non-linear
redshift space two-point correlation function is given by 
\begin{equation}
\xi^s ( s_\parallel ,s_\perp) = \int \xi^s _L ( s_\parallel +\upsilon ,
s_\perp ) {\it f}(\upsilon) d \upsilon 
\end{equation}
where the distribution function ${\it f }(\upsilon)$  is normalized to  
$\int^{\infty}_{-\infty} { \it f}(\upsilon) d \upsilon = 1 $. 
The authors who have invoked this model have generally adopted either a
Gaussian or else an exponential pairwise velocity distribution
function. In both cases, the distribution function has only one unknown
quantity $\sigma_R^2$ which is the velocity dispersion of the random
component of the pairwise peculiar velocity of the galaxies. In this
model, the observations of the anisotropies in $\xi^s$ can be used to
jointly determine the value of $\beta$ and $\sigma_R$. This has
recently been accomplished for the 2dFGRS where they find 
$\beta=0.49 \pm 0.09$ and $\sigma_R=506 \pm 52 \, {\rm km s^{-1}}$
\citep{haw}.  

An alternative approach is to attribute the deviations from the
linear predictions in $\xi^s$ to non-linear effects arising from the
coherent flows.  \citet{tayl}, \citet{fish96} and
\citet{hatton} have used the Zel'dovich approximation to analytically
study  the behaviour of the redshift-space power spectrum in the
translinear regime. They find that the results from the Zel'dovich 
approximation are in reasonable agreement with the predictions of
N-body simulations,  indicating that the coherent flows may be making
a significant contributions to the non-linear effects observed in
the redshift space two-point correlation function.  

In a different approach to studying the deviations in $\xi^s$ from the
linear predictions at scales where linear theory is known to be valid
in real space, \citet{bharad}  has considered the non-linear effects
introduced by the mapping from real space to redshift space. Under the
assumption that linear theory is valid in real space and that the
density fluctuations are a Gaussian random filed,   $\xi^s$ has been 
calculated taking into account all the non-linear effects that arise
due to the mapping from real to redshift space. It may be noted that 
the original calculation of \citet{kais} and \citet{ham} treats
the mapping from real to redshift space  to linear order only. 

In summary, at large scales where linear theory is known to be valid
in real space, the commonly used phenomenological model for $\xi^s$
attributes all the deviations from the linear predictions to the
effects of random motions on the mapping from real to redshift
space. On the other hand, \citet{bharad} calculated $\xi^s$
incorporating all the non-linear effects which arise in the mapping from
real to redshift space assuming that they are caused only by the
coherent flows. In all probability, the deviations  
from the linear predictions found in $\xi^s$  in the N-body
simulations and in actual redshift surveys is a consequence of
non-linear effects in the mapping from real space to redshift space
arising from both these effects namely, random motions and  coherent
flows. In this paper we consider models for the redshift space
distortions which combine both these effects, We compare the
predictions of these models with the commonly used phenomenological
model which has only the non-linear effects from random motions. 
We also compare all  these models with N-body
simulations and investigate which model best  fits the N-body results.
The different models are presented in 
Section 2 and the results of the comparison with N-body simulations
are presented in Section 3. 
The galaxy pairwise  velocity dispersion is a quantity which crops up
in any discussion of the effects of redshift space distortions on the
two-point correlation function. This quantity is very interesting in
its own right and it has received a considerable amount of attention
(\citealt{dav83}, \citealt{bin},\citealt{mojingbo}, \citealt{brain}, 
\citealt{somr},\citealt{bharad97},\citealt{mojing}, \citealt{landy}, 
\citealt{rat}, \citealt{striker}, \citealt{jingbo},\citealt{jingbor},
\citealt{bharad}, \citealt{sheth2}, \citealt{pop}
). This quantity has been observationally determined for different
redshift surveys(eg. \citealt{jingmobo}, LCRS ; \citealt{jevi}, SDSS ;
\citealt{haw} ,2dFGRS ).  The models we use for the redshift space
distortion also make definite  predictions for the pair-wise velocity
dispersion.   In Section 4 we calculate the pair-wise velocity
dispersion predicted by the different models and compare these with
the pair-wise velocity dispersion determined directly from the N-body 
simulations. 

In Section 5 we discuss our results and present conclusions. 

We would also like to point out that the models which we have
considered for $\xi^s$  are very similar in spirit to those considered
by  \citet{mats}, \citet{regos} and \citet{fish95}.

\section{Modeling $\xi^s$}

The two-point statistics of the galaxy distribution in real space is
completely quantified
 by the phase space distribution function
$\rho_2(\r,\v_1,\v_2)$ which gives the  probability density of finding
a  galaxy pair at a separation $\r$, one member of the pair having
peculiar velocity $\v_1$ and the other $\v_2$. The redshift space
two-point phase space distribution function  $\rho^s_2(\s,\v_1,\v_2)$
is 
related to its real space counterpart through
\begin{equation}
\rho^s_2(\s,\v_1,\v_2)=\rho_2(\s - \n \, U, \v_1,\v_2) 
\label{eq:b1}
\end{equation}
where we have assumed the plane parallel approximation, and the units
are chosen such that $H_0=1$.  
Here $U=\n \cdot (\v_2-\v_1)$ is the line of sight component of the
relative peculiar velocity of the galaxy pair.  Integrating out the
peculiar velocities gives us the redshift space two-point correlation
function 
\begin{equation}
1+\xi^s(\s)=\int \rho^s_2(\s,\v_1,\v_2) \, d^3 v_1 \, \, d^3 v_2
\label{eq:b2}
\end{equation}

We next introduce the key assumption in the model, the assumption
being that the peculiar velocity $\v$ of any galaxy can be written as a
sum of two parts $\v=\v_C+\v_R$, where $\v_C$ arises from large-scale 
coherent flows into overdense regions and out of underdense
regions, and $\v_R$ is a random part arising from galaxy motions in
virialized clusters and other non-linear regions. The large-scale
coherent flows are correlated with the density fluctuations which
produce the flows, and the two are assumed to be related through 
linear theory. The two-point statistics of the coherent flow is
quantified through the distribution function
$\rho_{2C}(\r,\v_{1C},\v_{2C})$ which is defined in exactly the same
way as $\rho_2$ the only difference being that $\rho_{2C}$ refers to
only the part of the peculiar velocities which arises from the
coherent flows. The statistical properties of the random part of the
peculiar velocity are assumed to be isotropic and independent of the 
galaxy's location. Its' joint probability density can be written as  
$\rho_{2R}(\v_{1R},\v_{2R}) = g([\v_{1R}]_x)g([\v_{1R}]_y) \,...\,
g([\v_{2R}]_z)$  where $[\v_{1R}]_x, [\v_{1R}]_y$ etc. refer to the
different Cartesian components of $\v_{1R}$ and $\v_{2R}$, and  
$g(v_R)$ is the distribution function for a single component of  the
random part of a galaxy's peculiar velocity. The joint distribution of
$\v_1=\v_{1C}+\v_{1R}$  and $\v_2=\v_{2C}+\v_{2R}$  can be expressed
in terms of the distribution functions for $\v_{1C},\v_{2C}$ and 
$\v_{1R},\v_{2R}$ as 
\begin{equation}
\rho_2(\r,\v_1,\v_2)=\int d^3v_{1R} \, d^3v_{2R} \, 
\rho_{2C}(\r,\v_1-\v_{1R},\v_2-\v_{2R}) \, 
\rho_{2R}(\v_{1R},\v_{2R})
\label{eq:b3}
\end{equation}
Using this in equations (\ref{eq:b1}) and (\ref{eq:b2}) to calculate the
redshift space two-point correlation function we have 
\begin{equation}
1+\xi^s(\s)=\int d u_{1} \, d u_{2} \, \left[ \int d^3~v_1 \,
  d^3~v_2 \,  \rho_{2C}(\s-\n (U+u_2-u_1),\v_1, \v_{2}) \, \right] \, 
g(u_1)  \, g(u_2) 
\label{eq:b4}
\end{equation}
where $u_1$ and $u_2$ are the line of sight components of $\v_{1R}$
and $\v_{2R}$ respectively. The term in the square 
brackets $[..]$ in equation (\ref{eq:b4}) can, on comparison with
equations (\ref{eq:b1}) and (\ref{eq:b2}), be identified as the
redshift space two-point correlation function if only the effects of
the coherent flows are taken into account 
\begin{equation}
1+\xi^s_C(\s)=\int d^3v_1  d^3v_2 \rho_{2C}(\s-\n U, \v_1,\v_2) 
\label{eq:b5}
\end{equation}
and $\xi^s$ can be expressed as 
\begin{equation}
\xi^s(\s)=\int du_1 \, du_2 \, \xi^s_{C}(\s-\n(u_2-u_1)) \, g(u_1) \,
g(u_2) \,. 
\label{eq:b6}
\end{equation}
To summarize, we start from the assumption that the galaxy peculiar
velocities have two parts, one from the coherent flows and the other
from random motions.  We show that the redshift space
correlation function $\xi^s$ is  $\xi^s_C$, which has only the
effect of the coherent flows, convolved along the line of sight with
the one-dimensional distribution function of the random part of the
galaxy's peculiar velocity, there being two convolutions, one for each
galaxy in the pair.  

The fact that only the relative peculiar velocity $v=u_2-u_1$ between 
the two galaxies appears in equation (\ref{eq:b6})   allows us to
simplify it a little further. Equation (\ref{eq:b6}) can be expressed
it in terms of the self-convolution of  $g(v_R)$  
\begin{equation}
f(v)=\int \, g(v-u) \, g(u) \, du 
\label{eq:b7} \, .
\end{equation}
The function $f(v)$ may be interpreted as the distribution function
for the line of sight component of the random part of the relative
peculiar velocity $v=u_2-u_1$ which is also called the pairwise
velocity. Using this, we finally obtain $\xi^s$ in terms of $\xi^s_C$
as 
\begin{equation}
\xi^s(s_{\|},s_\bot) = \int dv \, \xi^s_{C}(s_{\|}+v,s_\bot) \, f(v) \,
\label{eq:b8}
\end{equation}

We now shift our attention to $\xi^s_C$, the redshift space two-point
correlation function if only the coherent flows are taken into
account. As mentioned earlier, we assume that we are working at large
scales where linear theory holds in real space and the density
fluctuations are a Gaussian random field. Expanding $\rho_{2C}(\s-\n
U, \v_1,\v_2)$ in equation (\ref{eq:b5}) in a Taylor series in the
relative peculiar velocity $U$ of the coherent flow  we have 
\begin{equation}
1+\xi^s_C(\s)=\sum_{n=0}^{\infty} \frac{(-1)^n}{n!} 
\left(\frac{\partial}{\partial s_{\|}}\right)^n  
\int d^3v_1 d^3v_2 \, U^n \, \rho_{2C}(\s,\v_1,\v_2)  
\label{eq:b9}
\end{equation}

Retaining only the terms to  order $n=2$ we have 
\begin{equation}
\xi^s_L\ss=\xi(s) - \frac{\partial}{\partial \st} \vp\ss 
+ \frac{1}{2} \frac{\partial^{2}}{\partial \st^{2}} 
\sig\ss 
\label{eq:b10}
\end{equation}
where $\xi$ is the galaxy two-point correlation function in real
space, $\vp$ is the line of sight component of the mean  relative
velocity  between the galaxy pair (also called mean pairwise velocity) 
\begin{eqnarray}
\vp\ss&=&\int d^3v_1 d^3v_2 \, U \, \rho_{2C}(\s,\v_1,\v_2) \\
&=&  -\frac{2}{3} \st \, \beta \,  \xib_2(s) 
\label{eq:b11}
\end{eqnarray}
and $\sig$ is the mean square of the line of sight component of
the relative peculiar velocity (also called the pairwise velocity
dispersion) 
\begin{eqnarray}
\sig \ss&=&\int d^3v_1 d^3v_2 \, U^2 \, \rho_{2C}(\s,\v_1,\v_2) \\
&=& \beta^2 \left[ \frac{s^2}{3} \xib_1(s) - \frac{s^2_{\perp}}{3} 
\xib_2(s) + \frac{(s^2 - 3 s^2_{\parallel})}{15} \xib_4(s) \right]
\label{eq:b12}
\end{eqnarray}

Equation (\ref{eq:b10}), combined  with equations (\ref{eq:b11}) and 
(\ref{eq:b12}),  
is exactly the same as the linear redshift space two-point correlation
function calculated by \citet{ham}. Decomposing the angular dependence
of equation (\ref{eq:b10}) into Legendre polynomials one recovers
exactly the same angular moments as equation (\ref{eq:a2}),
(\ref{eq:a3}) and (\ref{eq:a4}), and  the odd moments and all even
moments beyond $l=4$ are zero. Using $\xi^s_L$ as given by equation
(\ref{eq:b10}) in equation (\ref{eq:b8}) corresponds to the
phenomenological model discussed earlier for the non-linear effects in
$\xi^s$, and this 
is one of the models which we shall be considering in the paper. 

Going back to equation (\ref{eq:b9}) for $\xi^s_C$, it is possible to
exactly sum up the whole series keeping all powers of $U$
\citep{bharad}. All the non-linear effects which arise due to the
mapping  from real space to redshift space are taken into account in
this calculation, and the resulting  redshift space two-point
correlation function is given by 

\begin{eqnarray}
1+\xi^{s}_{\LL}\ss &=&  \int   d \spt \,  G(\spt,\sigm\spp) \times  
\label{eq:a20}
\\
&\times&\left[\xi_{r}\spp + \left(1  - \frac{\spt \vp\spp} {2 \sig\spp}
\right)^2 - \frac{\vp^2\spp}{4 \sig\spp} \right] 
 \, \,. \nonumber 
\end{eqnarray}
where we use 
\begin{equation}
G(x,a)=\frac{1}{\sqrt{2 \pi} a} \exp[- \frac{x^2}{2 a^2}]
\end{equation}
to represent a normalized Gaussian distribution. 

We now have two different possibilities, $\xi^s_L$ or $\xi^s_{LL}$,
which we can use for $\xi^s_C$ in equation (\ref{eq:b8}) to calculate
the full redshift space  two-point correlation function  $\xi^s$.  
The function $\xi^s_C$ has only the effect of the coherent flows and
it has to be convolved with  $f(v)$, the one dimensional distribution
function for the random part of the pairwise velocity,  to
calculate $\xi^s$.  In this paper we have tried out four different
models which correspond to for different choices for $\xi^s_C$ and
$f(v)$. These are listed in Table I.

\begin{center}
\begin{table}{Table I.}\\
\begin{tabular}{|c|c|c|}
\hline
Model & $\xi^s_C$ & $f(v)$ \\
\hline
A & $\xi^s_L$ & $\frac{1}{\sqrt{2}\sigma_R}\exp \left( \frac{-
  \sqrt{2} \mid v  \mid} {\sigma_R}\right)$ \\ \\
B & $\xi^s_{\LL}$ & $\frac{1}{\sqrt{2}\sigma_R}\exp \left( \frac{-
  \sqrt{2} \mid v  \mid} {\sigma_R}\right)$ \\ \\
C & $\xi^s_{\LL}$ & $\frac{1}{\sigma_R^2}\exp\left(\frac{- 2 \mid v
    \mid}{\sigma_R} \right) \left(\frac{\sigma_R}{2} +\mid v \mid
\right)$ \\ \\
D & $\xi^s_{\LL}$ & $ \frac{1}{\sqrt{2\pi}\sigma_R} \exp
\left(  \frac{-v^2}{2 \sigma_R^2} \right) $ \\ \\
\hline
\end{tabular}
\end{table}
\end{center}

To highlight the salient features of the four models, Model A
uses $\xi^s_L$ for $\xi^s_C$ and an exponential for $f(v)$. This  is
the  phenomenological model discussed earlier. This model has been
used extensively by different people when analyzing both N-body
simulations and actual redshift surveys.  Models B, C and D all use  
$\xi^s_{\LL}$. The difference between these models is in the choice of
$f(v)$. Model B 
uses an exponential form for $f(v)$ and model D a Gaussian. Model C
corresponds to a situation where the one dimensional distribution
function for the random part of the galaxy peculiar velocity $g(u)$ is
assumed to be an exponential. The function $f(v)$ is now  the
convolution of two exponentials. All the models for $f(v)$ have only
one free parameter,  $\sigma^2_R$ which may be interpreted as the
pairwise velocity dispersion of the random part of the peculiar
velocity. 

In the next section we test the predictions of these models against
the results of N-body simulations. 

\section{Results for $\xi^s$}

In this section we calculate $\xi^s$ for the four models discussed
earlier and compare the results  against the predictions of N-body 
simulations. 
\subsection{The N-body Simulations.}
We have used a Particle-Mesh (PM) N-body code to simulate
the present distribution of dark matter in a comoving  region $[179.2
\, h^{-1} \, {\rm Mpc}]^3$. The simulations were run
using  $256^3$ grid points at $0.7\,h^{-1}\, {\rm Mpc} $ spacing
with $128^3$ particles for a ${\rm \Lambda CDM}$  cosmological model
with $\Omega_{m0}=0.3$, $\Omega_{\Lambda0}=0.7$ and $h=0.7$. We have
used a COBE normalized power spectrum with the shape parameter
$\Gamma=0.2$ for which $\sigma_8=1.03$. 

The low resolution N-body simulation used here is adequate for
studying the deviations from the predictions of linear theory in
redshift space on scales where the real space density
fluctuations are well described by linear theory. We have restricted
our analysis to scales larger than $5 \, h^{-1} \, {\rm Mpc}$, though
strictly speaking we would expect linear theory to be valid at scales
larger than something like $8 \, h^{-1} \, {\rm Mpc}$. To test that
our low resolution simulations are not missing out any crucial feature
either in real space or in redshift space, we have compared the
results of our N-body simulations with the Virgo simulations
\citep{jenk} which  have a higher resolution and a slightly different
normalization with $\sigma_8=0.9$. We find that on the length-scales
studied here, the results of our simulation are consistent with the
Virgo simulation both in real and redshift space. We show the  results
from the Virgo simulation alongside with those from our N-body
simulation.    Our N-body simulation was run for five independent
realizations of the initial conditions. 

Assuming that galaxies trace mass, $10^5$ dark matter particles were 
chosen at random from the simulation volume and the entire analysis was
carried out using these. The particle distribution in real space was
taken over to redshift space in the plane parallel approximation.  We
determined  the two-point correlation  
function for the particle distribution both in real and in redshift
space.  The angular dependence of the redshift space two-point
correlation function was decomposed into Legendre polynomials, and the
anisotropy in $\xi^s$ quantified through the  ratios
$\xi_2(s)/\xi_0(s)$ and $\xi_4(s)/\xi_0(s)$.  We also estimated 
the ratio Q(s) (eq. \ref{eq:a6}) which is somewhat different from
${\xi_2(s)}/{\xi_0(s)}$ in the sense  that it  uses an integrated
clustering measure instead of $\xi_0(s)$. This has the  advantage 
that in the linear theory of redshift distortion the value of
$Q(s)$ is expected to be a constant irrespective of the shape of the
real space correlation $\xi(s)$.  
Our simulations  have $\Omega_{m0}=0.3$ and $b=1$ which corresponds to
$\beta=0.49$,  and we expect $Q(s)=0.57$.

The average and the $1 \, \sigma$ errorbars for $\xi_0$,
$\xi_2/\xi_0$, $Q$ and $\xi_4/\xi_0$   were calculated using the five
realizations of our 
N-body simulations and the results are shown in Figures 1 to 4
respectively.  The points to note are 
\begin{itemize}
\item[(a.)] The results of our simulation are consistent with those of
  the Virgo simulation which are also shown in the figures
\item[(b.)] We see substantial deviations from the predictions of
  linear theory in redshift space on scales where it is known to hold
  in real space. This is best seen in the behaviour of $Q(s)$ which is
  supposed to be a constant with value $0.57$. We find that the value
  of $Q$ is much below this even at scales as large as $20 \, h^{-1}
  {\rm Mpc}$. The values of $Q$ increases gradually toward
  the linear prediction all the way to length-scales as large as $30-40
\,   h^{-1} \, {\rm Mpc}$ where it finally appears to saturate at the
  linear prediction. 
\item[(c.)] The errorbars increas with increasing pair separation and
  they are quite large beyond $25 \, h^{-1} \, {\rm Mpc}$. We have
  tried using a larger number of particles to estimate $\xi^s$ but
  this does not reduce the errorbars leading to the conclusion that
  the we are limited by the cosmic-variance arising from the finite
  size of our simulation and not by Poisson noise. Larger simulations
  will be required to make more accurate predictions for the nature of
  the redshift space anisotropies. 
\end{itemize}
\subsection{Fitting the models to N-body simulations}

All the models require the real space quantities $\xi(s)$, 
$V_p(s_{\perp},s_{\parallel})$  and $\sigma_P^2(s_{\perp}, 
s_{\parallel})$ as inputs to calculate $\xi^s$ in redshift space. 
We use  $\xi$, the  real space correlation function averaged over  five
realizations of the N-body simulation,   and its moments to calculate 
$V_p(s_{\perp},s_{\parallel})$ and
$\sigma_P^2(s_{\perp},s_{\parallel})$ using  equations (\ref{eq:b11}) 
and  (\ref{eq:b12}) respectively. Again, calculating $\xi^s$ using any
of our models requires us to specify  $\beta$ and $\sigma_R$. 
We have used $\beta=0.49$ which is the value corresponding to the
simulation parameters,  and we treat $\sigma_R$ as a free parameter
which we vary to obtain the best fit to the N-body results. 
For each model we fitted  the model predictions for 
$\xi_2(s)/\xi_0(s)$ and $Q(s)$ to the N-body results  using a $\chi^2$
minimization with $\sigma_R$ as the fitting  parameter. There are good
reason to believe that linear theory will  not hold for $s<8 \,h^{-1}
\,{\rm Mpc}$  and the fit was restricted to the region $8 \le s \le 40
\, h^{-1} \, {\rm Mpc}$. To check if
the models also work on length-scales which are  mildly non-linear in
real space, we have also carried out the fitting over the range 
$5 \le s \le 40\, h^{-1} \, {\rm Mpc}$.

We find that for  both $\xi_2(s)/\xi_0(s)$ and $Q(s)$, the
value of $\chi^2$ is minimized at nearly the same value of $\sigma_R$,
and so we quote the values only for $Q(s)$.  The best fit values of
$\sigma_R$  and the corresponding values of ${\chi_{min}^2}$ per
degree of freedom ${\nu}$ are  listed in Table II. 
The  model predictions  at the value of $\sigma_R$ which gives the
best fit in the interval $8 \le s \le 40 \, h^{-1} \, {\rm Mpc}$ are
shown  along with  the results of our N-body simulations in
figures 1 to 4. 

We find that all the models give a very good fit to the monopole
(Figure 1), and the best fit predictions of  the different models are 
indistinguishable from one another. Considering next the anisotropies
in $\xi^s$ (Figures 2 and 3) over the length-scales $8 \le s \le 40 \,
h^{-1} \, {\rm   Mpc}$, we find 
that all the models give a reasonably good fit.   Model C has the
smallest best fit $\chi^2/\nu$, and   Models B,  A and D follow in
order of 
increasing $\chi^2/\nu$. The values are $< 1$ for all the models,
indicating that all of them give acceptable fits. 
 It should be noted that the best fit values of
$\sigma_R$ vary considerably across the different models, and Model A
 predicts a value  considerably larger than the other models. 
Shifting our attention to the fits over the length-scales 
$5 \le s \le 40 \, h^{-1} \, {\rm Mpc}$ we find that model B gives the
lowest value of the best fit $\chi^2/\nu$, followed by Models A, C and 
D. All the models, except model D, have best fit $\chi^2/\nu$ below
unity and hence give acceptable fits.  
Interestingly, the acceptable models A, B and C    seem to
work better than 
one would expect  given the fact that the  length-scales $\sim  5 \,
h^{-1} \, {\rm Mpc}$ would be mildly non-linear in real
space.  Model D shows considerable
deviations from the N-body results at length-scales  $5
\le s \le 8 \, h^{-1} \, {\rm Mpc}$. Here again, the best fit values
of $\sigma_R$ show considerable variations across the models. Also,
for the same 
model,  the best fit $\sigma_R$ changes considerably when the fitting is
done over length-scales  $5 \le s \le 40 \, h^{-1} \, {\rm Mpc}$
instead of $8\le s \le 40 \, h^{-1} \, {\rm Mpc}$. This is
particularly noticeable for Model D where best fit $\sigma_R$
decreases by $\sim 25 \%$  when the fiting is extended to  smaller
length-scales. This change is $\sim 10 \%$ for
Models A and C, and $ \sim 5 \%$ for Model B.  It should also be
noted that for Model A, the best fit $\sigma_R$ increases when the
fitting is extended to smaller length-scales,  whereas the effect is 
opposite in all the other models. 

We now turn our attention to the hexadecapole ratio $\xi_4/\xi_0$
(Figure 4). 
Here again, for all the models  we use the values of $\sigma_R$ for
which the model predictions for $Q(s)$ give the best fit to the N-body
results. The ratio $\xi_4/\xi_0$ calculated with these values of
$\sigma_R$ are shown in Figure 4. We find that in the range $10 - 22
\, h^{-1} {\rm Mpc}$   the predictions of all the models fall below the
N-body results, These deviations are within the $1 \sigma$ errorbars
and larger simulations are required before we can be really sure of
the statistical significance of this effect. We have also tried
fitting our models to the N-body 
results using a $\chi^2$ minimization for $\xi_4/\xi_0$ with
$\sigma_R$ as the free parameter. The best fit $\sigma_R$ obtained
this way are quite different from those obtained by fitting 
$\xi_2/\xi_0$ and $Q(s)$ and we do not report these values here. 
This discrepancy may be indicating the inability of these models to
adequately describe the hexadecapole $\xi_4$, but
further studies using larger N-body simulations with smaller errorbars
are required to reach a definite conclusion.

\begin{center}
\begin{table}{Table II.}\\
\begin{tabular}{|c|c|c|c|c|}
\hline
Model & $\sigma_R(\rm km/sec)$ & $\chi_{min}^2/\nu$ 
& $\sigma_R(\rm km/sec)$ & $\chi_{min}^2/\nu$\\
\hline
& $8 \le s \le 40 \, h^{-1} {\rm  Mpc} $ & & $5 \le s \le 40 \, h^{-1}
  {\rm  Mpc} $ &  \\ 
\hline
A & $684$ & $0.055$ & $760$ & $0.51$  \\
B & $564$ & $0.054$ & $540$ & $0.19$   \\
C & $520$ & $0.050$ & $452$ & $0.86$  \\
D & $489$ & $0.114$ & $367$ & $2.20$  \\
\hline
\end{tabular}
\end{table}
\end{center}

\begin{figure}
\includegraphics[width=100mm]{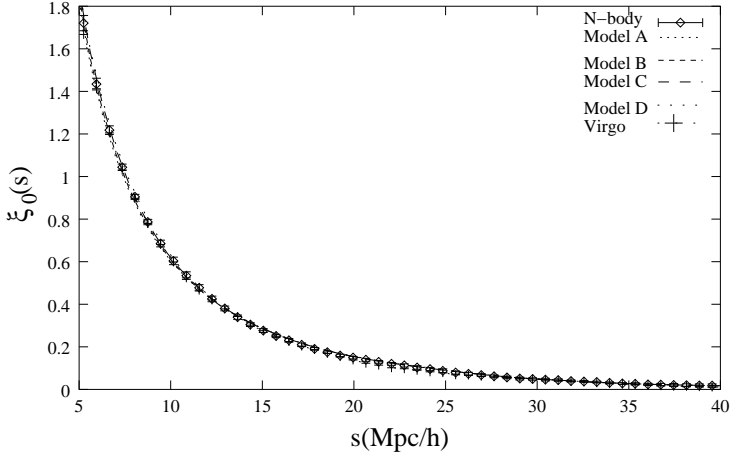}
\caption{This shows the monopole
  $\xi^s_0$ as determined from our  N-body simulations and the Virgo
  simulation. The normalization of the power spectrum used in the
  Virgo simulation is   slightly different from the one used by us
  (Section 3.1), and the results from the Virgo simulation have been
appropriately scaled to compensate for this.    
The figure also shows the predictions of the four models
  considered here for the value of $\sigma_R$ (Table II) which gives
  the best fit   to $Q(s)$ in the interval $8 \le s \le 40 \, h^{-1}
  {\rm Mpc}$. The outcome of our simulations, the Virgo simulation
  and the best fit predictions of all four models are
  indistinguishable   from one another.}
\label{fig:a1}
\end{figure}

\begin{figure}
\includegraphics[width=100mm]{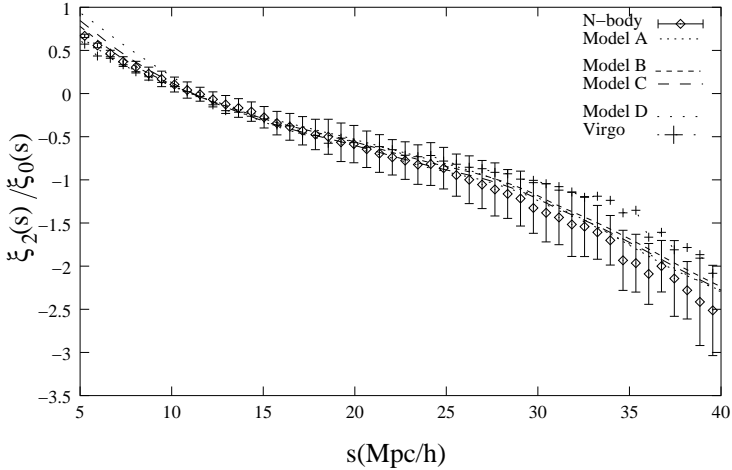}
\caption{This shows the ratio 
  $\xi^s_2/\xi^s_0$ as determined from our  N-body simulations and the
  Virgo   simulation. It also shows the predictions of the four
  models   considered here for the  value of $\sigma_R$   which gives
  the best fit   in the interval $8 \le   s \le 40 \, h^{-1}   {\rm
  Mpc}$ (Table II).} 
\label{fig:a2}
\end{figure}

\begin{figure}
\includegraphics[width=100mm]{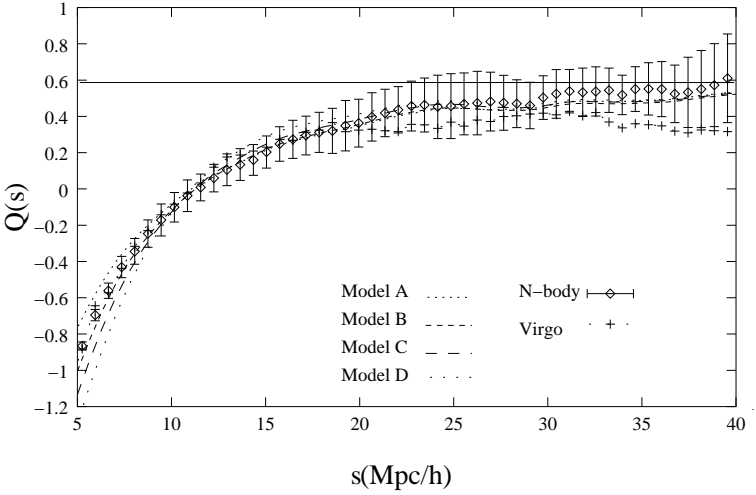}
\caption{This shows $Q(s)$ as determined from our  N-body
  simulations and the   Virgo   simulations. It also shows the
  predictions of the four   models   considered here for the  value of
  $\sigma_R$   which gives   the best fit   in the interval $8 \le   s
  \le 40 \, h^{-1}   {\rm   Mpc}$ (Table II). The horizontal line at
  $Q(s)=0.57$ is the constant value predicted by the linear theory of
  redshift distortions.}
\label{fig:a3}
\end{figure}

\begin{figure}
\includegraphics[width=100mm]{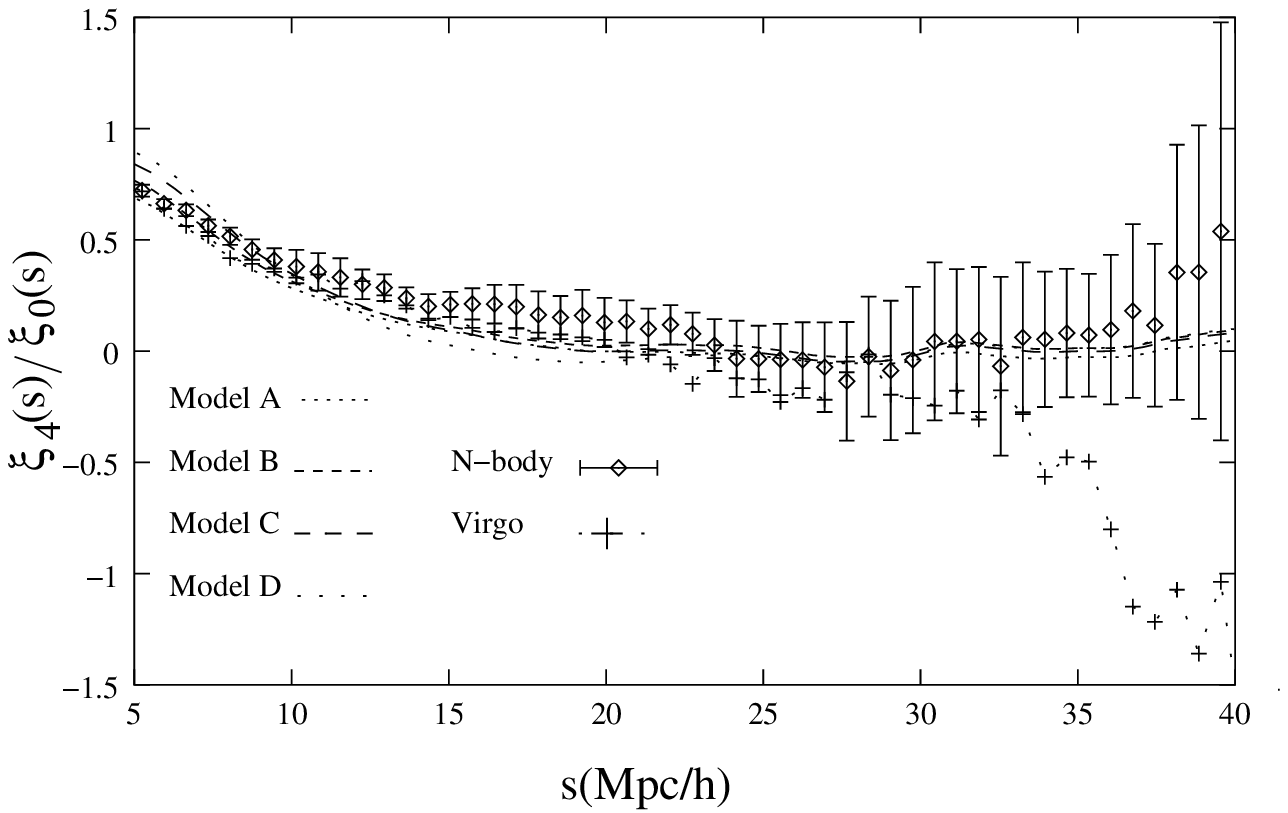}
\caption{This shows the ratio 
  $\xi^s_4/\xi^s_0$ as determined from our  N-body simulations and the
  Virgo   simulations. It also shows the predictions of the four
  models   considered here. It should be noted that model predictions
  are for  the  value of $\sigma_R$   which gives
  the best fit  to $Q(r)$ and not $\xi^s_4/\xi^s_0$ in the interval $8
  \le   s \le 40 \, h^{-1}   {\rm   Mpc}$ (Table II).} 
\label{fig:a4}
\end{figure}

\section{The Pairwise velocity Dispersions}

The pairwise velocity dispersion is an important statistical quantity
which sheds light on the clustering of matter in the universe. 
There are several approaches to determine the pairwise velocity
dispersion on small scales from observations, for example,  using  the
cosmic virial 
theorem (\citealt{peeb}, \citealt{sutjin}, \citealt{pop}) or by
modeling the distortions in the redshift-space correlation function
(eg. \citealt{davgel}, \citealt{dav83}, \citealt{bin},
\citealt{mojingbo}, \citealt{jingmobo}, \citealt{jingbo},
\citealt{landy}, \citealt{rat}, \citealt{jevi}, \citealt{haw}).   

Our interest lies in the fact that the models which we have used  to
fit $\xi^s$ also make definite predictions for the pairwise
velocity dispersion at large scales where we expect linear theory to
hold. The pairwise velocity dispersion $\sigma^2_{ij}$, a  symmetric
rank two tensor, is defined as the second moment of the relative
velocities of galaxy pairs and its value can be calculated from the
distribution function $\rho_2(\r,\v_1,\v_2)$ as
\begin{equation}
\sigma^2_{ij}(\r) =\int (\v_2-\v_1)_i  \, (\v_2-\v_1)_j \,
\rho_2(\r,\v_1,\v_2) d^3 v_1 d^3 v_2 / [1+\xi(r)] 
\label{eq:c1}
\end{equation}
where $i,j$ refer to different Cartesian components. 
Our work is restricted to large scales where linear theory holds in
real space and  we use $1+\xi(r) \approx 1$. 
The most general form for  $\sigma^2_{ij}(\r)$ which is consistent with
statistical homogeneity and isotropy is 
\begin{equation}
\sigma^2_{ij}(\r)=\sigma^2_{\perp}(r) \delta_{ij}+
  [\sigma^2_{\parallel}(r) -   \sigma^2_{\perp}(r)] (\r_i \r_j/r^2) \,.
\label{eq:c2}
\end{equation}
Here $\sigma^2_{\perp}(r)$ is the pairwise velocity dispersion for the 
velocity component perpendicular to the pair
 separation $\r$ and  $\sigma^2_{\parallel}(r)$ is the dispersion for
 the velocity component parallel to $\r$. The behaviour of
 $\sigma^2_{ij}(r)$ is  completely specified through  these two
 components $\sigma^2_{\perp}(r)$ and $\sigma^2_{\parallel}(r)$. 
 We next recollect the fundamental assumption
underlying  all the models which we have considered in the previous
 section {\it ie.} the 
peculiar velocity of any galaxy has two parts, one arising from
coherent flows and another from random motions. Under this assumption
the two-point distribution function $\rho_2$ is the convolution of two
distribution functions (eq. \ref{eq:b3})  one describing the two-point
statistics of the coherent flow and another for the random motions.  
Using this in (equation \ref{eq:c1})  to calculate $\sigma^2_{ij}(r)$
gives us  
\begin{equation}      
\sigma^2_\parallel(r)=\sigma^2_{\parallel C}(r)+\sigma^2_R
\label{eq:c3}
\end{equation}
\begin{equation}
\sigma^2_\perp(r)=\sigma^2_{\perp C}(r)+\sigma^2_R
\label{eq:c4}
\end{equation}
 for all the models.  Here $\sigma^2_R$ is the
isotropic  contribution from random
 motions,  and $\sigma^2_{\parallel  C}(r)$ and $\sigma^2_{\perp C}(r)$
 are  the contributions from coherent flows.

Proceeding in exactly the same way as when using the models to fit
 $\xi^s$, we  assume that the coherent flows are related to the
 density fluctuations through linear theory {\it ie.} $\sigma^2_{\perp
 C}=\sigma^2_{\perp L}$ and  $\sigma^2_{\parallel
 C}=\sigma^2_{\parallel L}$.  This allows us to express 
$\sigma^2_{\parallel C}(r)$ and $\sigma^2_{\perp C}(r)$  in terms of
 the moments of the real space two-point  correlation  function
 \citep{bharad} as    
\begin{equation}
\sigma^2_{\parallel L}(r)=\beta^2  \, r^2 \,
  [\frac{1}{3}\bar\xi_1(r)-\frac{2}{15}\bar\xi_4(r)] 
\label{eq:c5}
\end{equation}

\begin{equation} 
\sigma^2_{\perp L}(r)=\beta^2 \,  r^2 \, [\frac{1}{3}\bar\xi_1(r)
  - 
  \frac{1}{2}\bar\xi_2(r) +    \frac{1}{15} \bar\xi_4(r)] \,. 
\label{eq:c6}
\end{equation}
In calculating $\sigma^2_{\parallel L}$ and $\sigma^2_{\perp L}$
we have used  the  average real space two-point correlation
function and its moments determined from our N-body simulations. 
In addition to $\sigma^2_{\parallel L}$ and $\sigma^2_{\perp L}$,  all 
the models  considered in  this paper also need the value of 
$\sigma_R$ as an input to calculate $\sigma^2_{\parallel}$ and
$\sigma^2_{\perp}$.  
 In Section 3, for each model we have
determined the best fit value of $\sigma_R$ (Table II) for which the
model predictions for $Q(s)$  best match the N-body results in the
range $8\le s \le 40 \, h^{-1} \, {\rm   Mpc}$. 
 We have used these
values of $\sigma_R$ to calculate the pairwise velocity dispersion
predicted by each of these models. The two independent components of
the pairwise velocity dispersion ($\sigma^2_{\parallel}$ and 
$\sigma^2_{\perp}$)   were also determined directly from
N-body simulations and the results are shown in Figure 5 and 6.

\begin{figure}
\includegraphics[width=100mm]{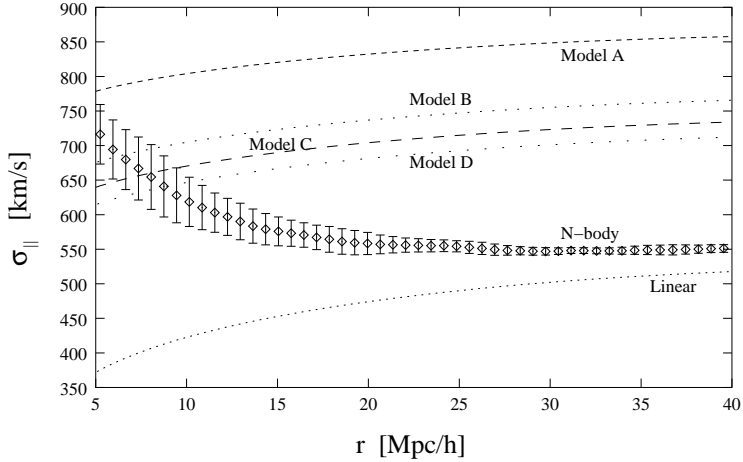}
\caption{ This shows $\sigma_{\parallel}$ as determined from  our
  N-body simulation,  along with the predictions of linear theory
  (eq. \ref{eq:c5}) and all the models considered in Sections 3. The
  models differ from the linear predictions in that they also have a
  contribution from random motions  added in
  quadrature to the linear predictions (eq. \ref{eq:c3}).}  
\label{fig:a5}
\end{figure}

\begin{figure}
\includegraphics[width=100mm]{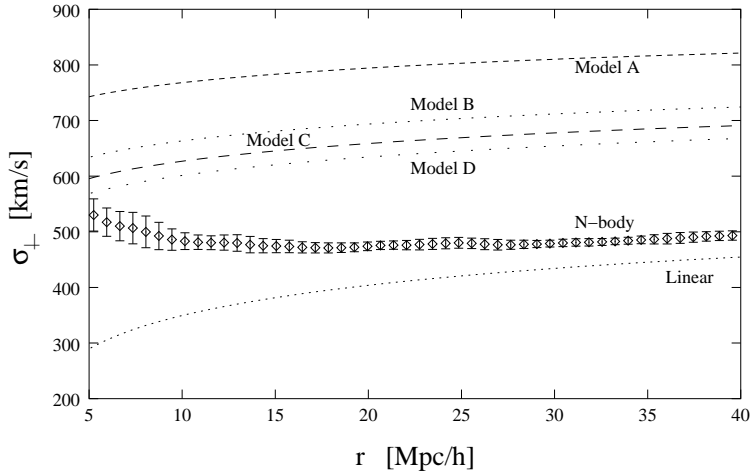}
\caption{ This shows $\sigma_{\perp}$ as determined from  our
  N-body simulation,  along with the predictions of linear theory
  (eq. \ref{eq:c6}) and all the models considered in Sections 3. The
  models differ from the linear predictions in that they also have an 
  contribution from random motions  added in
  quadrature to the linear predictions (eq. \ref{eq:c4}).}  
\label{fig:a6}
\end{figure}

We find  that $\sigma_{\parallel}$ and $\sigma_{\perp}$  determined
  from our N-body simulations decreases with increasing $r$ at
  length-scales $r \le 15 \, h^{-1} {\rm Mpc}$, after which it is
  more or less constant   with possibly a very slow variation with
  $r$.  Our N-body results are consistent with the high resolution
  simulations of  \citet{jenk}. 
 It is important to note that the variation of
  $\sigma^2_{ij}(r)$ with $r$ plays an important role in 
  redshift  space distortions. For example, at linear order
  (eq. \ref{eq:b10}) the redshift space two-point correlation function
  $\xi^s \ss$ depends explicitly on  $\frac{\partial^{2}}{\partial 
  \st^{2}}  \sig\ss$ which is the second derivative of the line of
  sight component of the pairwise  velocity dispersion. All the terms
  involving $\beta^2$ in  the expressions for the different angular
  moments of $\xi^s$  (eqs. \ref{eq:a3}, \ref{eq:a4} and  \ref{eq:a5})
  arise from  this.  The spatial variation of $\sigma^2_{ij}(r)$ also
  plays an important role in determining $\xi^s$ in equation
  (\ref{eq:a20})  where  all the   non-linear effects of the
  mapping from real to redshift space are   taken into account. 

Turning our attention to the model predictions, we first consider
$\sigma_{\parallel}$ and $\sigma_{\perp}$ calculated using only linear
theory (eqs. \ref{eq:c5} and \ref{eq:c6}) with the real space
correlations $\xi(r)$ and its moments determined from N-body
simulations. as inputs. We find that these fall short of the values of    
$\sigma_{\parallel}$ and $\sigma_{\perp}$ determined directly from
N-body simulations. Also, the $r$ dependence of $\sigma_{\parallel}$
and $\sigma_{\perp}$ are quite different, with the N-body results
decreasing and the linear predictions increasing with increasing $r$.   
At length-scales $r \ge 25 \, h^{-1} \, {\rm Mpc}$, the curves showing
linear theory and  the  N-body  results are approximately parallel,
with the linear predictions being approximately $50 \, {\rm km/s}$
below the N-body results. 

The model predictions differ from the linear theory predictions in
that they have a contribution from random motions $\sigma_R$ added in
quadrature to the linear predictions (eqs. \ref{eq:c3} and
\ref{eq:c4}). One might hope that the contribution from random motions
will compensate for the shortfall in the linear predictions relative
to the N-body simulations, and the predictions of the two will match
at least at length-scales $r \ge 25 \, h^{-1} \, {\rm Mpc}$ where the
two curves are parallel. The problem is that all the models predict
different values for $\sigma_R$, and the predicted values are too
large. Model A which has the  highest value of $\sigma_R$ fares the
worst with the predicted $\sigma_{\parallel}$ and $\sigma_{\perp}$
being  much larger than the N-body results.  The predictions of 
Models B, C and D are slightly closer to the N-body results, but they
are all still very significantly higher than the N-body results. In
summary $\sigma_{\parallel}$ and $\sigma_{\perp}$ predicted by all the
models are significantly in excess of the values determined directly
from N-body simulations. This indicates that there is a gap in our
understanding of what is really going on. 

The possibility of using the pairwise velocity dispersion as a tool for
distinguishing  between different cosmological models has been
controversial and this has been hotly debated in the  literature
(eg. \citealt{ostuto},  \citealt{cenker},  \citealt{couch},
\citealt{gelber}, \citealt{jur}, \citealt{branevil}, \citealt{brain},
\citealt{somr}).  An important fact that we should remember
while measuring the pairwise velocity dispersion from N-body
simulations is that it is a pair weighted statistic and is heavily
weighted by the densest regions present in the sample. These regions
naturally have the highest velocity dispersion and this tends to push
up the estimate. The statistic is strongly dependent on the presence
or absence of rich clusters within the sample (eg. \citealt{mojingbo},
\citealt{marj}, \citealt{mojing},  \citealt{somr},  \citealt{gujo},
 \citealt{hattons}). It has also been  
confirmed by several authors (eg. \citealt{sand}, \citealt{brown}, 
 \citealt{uilik}, \cite{striker}) that the  velocity field is very cold
outside the clusters. We note that these effects are not very crucial
in our work. This is because we have used exactly the same set of
particles drawn from our N-body simulations  to determine both $\xi^s$
and $\sigma^2_{ij}$, and we have been testing if the models which make
reasonably good predictions for $\xi^s$ are also successful in
correctly predicting $\sigma^2_{ij}$. We would expect this to be true
because the peculiar velocities which are quantified by the pairwise
velocity dispersion are also the cause of the redshift space
distortions. Surprisingly, we find that the model predictions for
$\sigma^2_{ij}$ are significantly in excess of $\sigma^2_{ij}$
determined directly from the simulations. 

\section{Discussion and Conclusions.}

The galaxy two-point correlation function determined from redshift
surveys shows significant deviations from the predictions of the
linear theory of redshift space distortion even on scales as large as
$20 - 30 \, h^{-1} {\rm Mpc}$ where linear theory is expected to be
valid on real  space. Any attempt to determine $\beta$ from redshift
surveys requires that these deviations  be properly modeled and taken
into account. Modeling redshift space distortions basically requires
a joint model for galaxy peculiar velocities and their correlations
with the galaxy clustering pattern.  Such models test our
understanding of the gravitational instability process by which the
large scale structures are believed to have formed.

We have considered four different models (details in Section 2) for
the  redshift space two-point correlation function $\xi^s$. All the
models are based on the assumption that galaxy peculiar velocities
may be decomposed into two parts, one arising from coherent flows and
another from random motions. It is also assumed that in real space the
coherent flows are well described by the linear theory of density
perturbation.  Deviations from the predictions of the
linear theory of redshift space distortion arise from  two
distinct causes  which affect the mapping from
real to redshift space (a.) non-linear effects  due to the coherent
flows  (b.) the random motions. Among the four models, Model A
does not incorporate the non-linear effects due to the coherent flows.  
It combines the predictions of the linear theory of redshift space
distortion (\citealt{kais}, \citealt{ham}) with the effect of the
random motions which is modeled through an exponential distribution
function for the pairwise velocity.  This is the popular
phenomenological model which has been widely applied to the analysis
of galaxy redshift surveys (eg. \citealt{haw}).   
Models B, C and D all take into account non-linear effects arising
from the coherent flows \citep{bharad}, and they differ from one
another in the choice of the distribution function for the random part
of the pairwise velocity.

All the models have only one free parameter, $\sigma_R$ which is the
one dimensional random pairwise velocity dispersion. For each  model
we have    determined the value of $\sigma_R$  for which the  model
predictions best fit the quadrupole anisotropy of $\xi^s$ determined
from N-body simulations. 
We find that Model C gives the lowest value of the best fit
$\chi^2/\nu$   over the range of length-scales
$8\le s \le 40 \, h^{-1} {\rm Mpc}$ where we expect linear theory to
be valid in real space. In this  model the distribution function for the
random part of a galaxy's peculiar velocity is modeled as an
exponential function.  It may be noted that the other three models
also give acceptable fits to the N-body results.

We find that three of the models (A, B and C) also give acceptable fits
over length-scales $5 \le s \le 40 \, h^{-1} {\rm Mpc}$ which includes
a small region where perturbations are expected to be mildly
non-linear. Model D where the distribution function $f(v)$
for the random part of  pairwise velocity  is a Gaussian fails
to give an acceptable fit. Model B where  $f(v)$ is an exponential
gives the lowest value of best fit $\chi^2/\nu$. 
It may be noted that though  the best fit value of $\chi^2/\nu$  for
model  A, the  commonly used phenomenological model, is around three times
larger than that for Model B,  it is not possible to draw a strong
statistical conclusion as to which model is superior. This is because
$\chi^2/\nu<1$  for Models A,B and C  and they all provide acceptable
fits. The present work is limited by the large statistical error-bars in
the quantities determined at large scales from N-body
simulations. These errors arise 
from the limited volume of the simulations (cosmic variance). It
should be possible to achieve lower $1-\sigma$ error-bars using larger
simulations whereby we could  distinguish between these
models at a higher level of statistical significance. We propose to
carry this out in the future.

Interestingly, the best fit value of $\sigma_R$ shows substantial
variations across the models. The best fit value of $\sigma_R$ is
substantially smaller in the  models which incorporate the
non-linear effects of the coherent flows (B, C and D) as compared to
Model A which does not include these effects. This indicates that
there are significant nonlinear effects in the mapping from real to
redshift space arising from the  coherent flows.  The commonly used
phenomenological model does not incorporate these effects and in this
model all deviations from the linear predictions are attributed to 
random motions. This leads to the pairwise velocity dispersion of the
random motion ($\sigma_R$) being  overestimated. For eaxample,
\citet{haw} have used Model A to fit the redshift space two-point
correlation function  of the 2dFGRS to obtain the best fit value
$\sigma_R=506 \pm 52 \, {\rm km/s}$.  The findings of this paper show
that Models B and C would be equally successful in fitting the same
observation,  and the best fit value of $\sigma_R$ would be different
for each of these models.  This raises questions as to the
interpretation of $\sigma_R$ determined by this method.

Although the models are all reasonably successful in fitting the
quadrupole anisotropies of $\xi^s$, the model predictions for the
pairwise velocity dispersion  are much  larger than the values
determined 
directly from N-body simulations. Surprisingly, the predictions of
linear theory which has a contribution from only the coherent flows
and not the random motions are much closer to the N-body results as 
compared to the model predictions. At large scales the predictions of
linear theory, all the models and the N-body results are all very
similar. The linear theory predictions are slightly below the N-body
results, and one would expect that it would be possible to recover the
N-body results by combining the linear theory predictions  with the
contribution from random motions. Unfortunately, all the models appear
to be overestimating the contribution from random motions and the
model predictions are significantly in excess of the N-body results.   
Also, the predictions of Model A fare the worst in comparison to the
other models. A possible explanation why equations (\ref{eq:c3}) and
(\ref{eq:c4}) overpredicts  the pairwise velocity dispersion is that
the linear component of the peculiar velocity also makes a
contribution to the random motions. It is possible that  this
is already present in $\sigma_R$, and it  contributes more than its
due share to the pairwise velocity dispersion.  

In the linear theory of  redshift space distortions  the hexadecapole
anisotropy arises from the 
line of sight component of the pairwise velocity dispersion. The fact
that none of the four models considered here  give a very good
fit to the hexadecapole is probably related to the fact that the
models also do not predict the correct   pairwise velocity dispersion.  

We note that the assumption that galaxy peculiar velocities
can be decomposed into two parts,  one coherent and another random is
consistent with the halo model. The random part may be attributed
to motions inside the halo and the coherent part to the overall
motion of the halo. \citet{seljak}, \citet{white} and \citet{kang}
have calculated 
the galaxy power spectrum in redshift space using the halo model. 
\citet{sheth1} have calculated the pairwise velocity dispersion using
the halo model. 

In conclusion we note that the nonlinear effects in the mapping from
real to redshift space introduced by the coherent flows are
important. Models which incorporate these effects provide an equally
good   to the quadrupole anisotropies  of $\xi^s$ as compared to
models which are based on the linear theory of redshift distortion.
Unfortunately, 
none of these models make correct predictions for the pairwise
velocity dispersion. This indicates that there is a gap in our
understanding of the statistical properties of the peculiar velocities
and their effect on the redshift space two-point correlation function.  

\section*{Acknowledgments}
BP is supported by a junior research fellowship of the
Council of Scientific and Industrial Research (CSIR), India.
SB would like to thank Jasjeet Bagla for a helpful discussion on the
analysis of N-body simulations. SB would also like to acknowledge
financial support from the Govt. of India, Department of Science and
Technology (SP/S2/K-05/2001). Both authors would like to thank Jatush
V Sheth for his help in analyzing the Virgo simulations. 

The Virgo simulations used in  this paper were carried out by the
Virgo Supercomputing Consortium using computers based at Computing
Centre of the Max-Planck Society in Garching and at the Edinburgh
Parallel Computing Centre. The data are publicly available at
www.mpa-garching.mpg.de/NumCos


\begin{thebibliography}{99}
\bibitem [\protect\citeauthoryear{Bharadwaj}{2001}]{bharad}
Bharadwaj, S., 2001, MNRAS,327,577
\bibitem [\protect\citeauthoryear{Bharadwaj}{1997}]{bharad97}
Bharadwaj, S., 1997,ApJ,477,1 
\bibitem [\protect\citeauthoryear{Bean et al.}{1983}]{bin}
Bean, A.J.,Efstathiou G.P.,Ellis R.S.,Peterson B.A.,Shanks
T.,1983,\mnras,205,605
\bibitem [\protect\citeauthoryear{Brainerd et al.}{1994}]{brain}
Brainerd, T. G., Bromley B. C., Warren M. S., Zurek W. H., 1996,
  \apj, 464, L103  
\bibitem [\protect\citeauthoryear{Brainerd \& Villumsen}{1994}]{branevil}
Brainerd, T.G., \& Villumsen J.V. ,1994,\apj,436,528
\bibitem[\protect\citeauthoryear{Bromley, Warren \& Zurek}{1997}]{brom}
Bromley B.C., Warren M. S., Zurek W. H., 1997,  \apj, 475, 414   
\bibitem[\protect\citeauthoryear{Ballinger,Peacock \& Heavens }{1996}]{bali}
Ballinger, W.E.,Peacock,J.A. and Heavens, A.F.,1996,MNRAS,282,877
\bibitem[\protect\citeauthoryear{Brown \& Peebles }{1987}]{brown}
Brown, M.E.,\& Peebles,P.J.E.,1987\apj,317,588
\bibitem[\protect\citeauthoryear{Cen \& Ostriker }{1992}]{cenker}
Cen, R. \& Ostriker, J.P.,1992,\apj,399,L113
\bibitem[\protect\citeauthoryear{Couchman \& Carlberg }{1992}]{couch}
Couchman, H.M.P., \& Carlberg, R.G.,1992,\apj,389,453
\bibitem[\protect\citeauthoryear{Davis \& Peebles}{1983}]{dav83}
Davis, M.~\& Peebles,P.~J.~E.\ 1983, \apj, 267, 465
\bibitem[\protect\citeauthoryear{Davis, Geller \& Huchra }{1983}]{davgel}
Davis, M.,Geller, M.J., \& Huchra, J. ,1978,\apj,221,1
\bibitem[\protect\citeauthoryear{Del Popolo}{2001}]{pop} 
Del Popolo, A.,2001,\mnras,326,667
\bibitem[\protect\citeauthoryear{Fisher et al.}{1994}]{fish94} Fisher,
K. B., Davis M., Strauss M. A., Yahil A., Huchra 
J. P., 1994, \mnras, 267, 927
\bibitem[\protect\citeauthoryear{Fisher}{1995}]{fish95} Fisher,
K. B.,  1995, \apj, 448, 494
\bibitem[\protect\citeauthoryear{Fisher \& Nusser}{1998}]{fish96}
Fisher, K.B. \& Nusser A. 1996, MNRAS, 279L, 1
\bibitem[\protect\citeauthoryear{Gelb \& Bertschinger}{1994}]{gelber}
Gelb, J.M., \& Bertschinger, E. ,1994,\apj,436,491
\bibitem[\protect\citeauthoryear{Guzzo et al.}{1997}]{gujo}
Guzzo, L.,Strauss, M.A.,Fisher, K.B.Giovanelli,R., \& Haynes,
M.P.,1997,\apj,489,37
\bibitem[\protect\citeauthoryear{Hamilton}{1992}]{ham} 
Hamilton, A. J. S. 1992, \apj, 385, L5  
\bibitem[\protect\citeauthoryear[{Hatton \& Cole}{1998}]{hatton}
  Hatton, S.~\& Cole, S.\  1998, \mnras, 296, 10
\bibitem[\protect\citeauthoryear[{Hatton \& Cole}{1999}]{hattons}
Hatton, S.~\& Cole, S.\ 1999, \mnras, 310, 1137
\bibitem[\protect\citeauthoryear{Hawkins et al.}{2002}]{haw}
Hawkins et al., 2003, MNRAS,346,78
\bibitem[\protect\citeauthoryear{Heavens \& Taylor}{1995}]{hevtayl}
Heavens, A.F. and Taylor, A.N.,1995,MNRAS,275,483
\bibitem[\protect\citeauthoryear{Jenkins et al.}{1998}]{jenk}
Jenkins et al., 1998,ApJ,499,20 
\bibitem[\protect\citeauthoryear{Jing, Mo \& Borner}{1998}]{jingmobo}
Jing, Y.P.,Mo, H.J.,Borner, G.,1998,\apj,494,1
\bibitem[\protect\citeauthoryear{Jing \& Borner}{1998}]{jingbo}
Jing, Y.P.,Borner, G.,1998,\apj,503,502
\bibitem[\protect\citeauthoryear{Jing \& Borner}{2001}]{jingbor}
Jing, Y.P.,Borner, G.,2001,\apj,547,545
\bibitem[\protect\citeauthoryear{Kaiser}{1987}]{kais}
Kaiser, N. 1987, \mnras, 227, 1
\bibitem[\protect\citeauthoryear{Kang et al.}{2002}]{kang}
Kang, X.,Jing, Y.P.,Mo, H.J.,Borner, G. \ 2002, \mnras, 336,892
\bibitem[\protect\citeauthoryear{Lahav et al.}{1991}]{lahav1} Lahav O.,
  Lilje, P. B., Primack, J. R. and Rees, M., 1991, MNRAS, 251, 128 
\bibitem[\protect\citeauthoryear{Landy, Szalay \& Broadhurst}{1998}]{landy}
Landy, S.D.,Szalay, A.S.,Broadhurst, T.,1998,\apj,494,L133
\bibitem[\protect\citeauthoryear{Matsubara}{1994}]{mats}
Matsubara, T., 1004, \apj, 424, 30
\bibitem[\protect\citeauthoryear{Mo,Jing \& Borner}{1993}]{mojingbo}
Mo, H.J.,Jing, Y.P. \& Borner, G.,1993,MNRAS,264,825
\bibitem[\protect\citeauthoryear{Mo,Jing \& Borner}{1997}]{mojing}
Mo, H.J.,Jing, Y.P. \& Borner, G.,1997,MNRAS,286,979
\bibitem[\protect\citeauthoryear{Marzke et al.}{1995}]{marj}
Marzke, R.O.,Geller, M.J.,da Costa, L.N. and Huchra,J.P.,1995,AJ,110,477
\bibitem[\protect\citeauthoryear{Ostriker \& Suto}{1990}]{ostuto}
Ostriker, J.P., \& Suto Y. ,1990\apj,348,378
\bibitem[\protect\citeauthoryear{Peacock \&
    Dodds}{1994}]{pcokanddodds} Peacock, J. A. \&  Dodds, S.J., 1994,
  \mnras, 267, 1020 
\bibitem[\protect\citeauthoryear{Peacock et al.}{2001}]{pcok}
Peacok, O.~et al.\ 2001, Nature, 410, 169 
\bibitem[\protect\citeauthoryear{Peebles}{1980}]{peeb} Peebles,
 P. J. E. 1990, The Large-Scale Structure of the Universe,
(Princeton: Princeton University Press) 
\bibitem[\protect\citeauthoryear{Ratcliffe et al.}{1998}]{rat}
Ratcliffe, A., Shanks, T., Parker, Q.A. and  Fong, R.,1998,MNRAS,296,191
\bibitem[\protect\citeauthoryear{Regos \& Szalay}{1995}]{regos}
 Regos, E.~\& Szalay, A.~S.\ 1995, \mnras, 272, 447 
\bibitem[\protect\citeauthoryear{Seljak}{2001}]{seljak} Seljak, U.\
  ,2001, \mnras, 325, 1359 
\bibitem[\protect\citeauthoryear{Sheth \& Diaferio}{2001}]{sheth1}
Sheth, R.~K.~\& Diaferio, A.\ 2001, \mnras, 322, 901
\bibitem[\protect\citeauthoryear{Sheth et al.}{2001}]{sheth2}
{{Sheth}, R.~K. and {Hui}, L. and {Diaferio}, A. and {Scoccimarro}, R.},
,2001,\mnras,325,1288
\bibitem[\protect\citeauthoryear{Sandage}{1986}]{sand}
Sandage,A. ,1986,\apj,307,1
\bibitem[\protect\citeauthoryear{Scoccimarro}{1986}]{scocci}
Scoccimarro, R. , 2004, astro-ph/0407214
\bibitem[\protect\citeauthoryear{Suto \& Suginohara}{1991}]{suto}Suto,
Y.,  Suginohara , T., 1991,\apjl, 370, L15    
\bibitem[\protect\citeauthoryear{Suto \& Suginohara}{1991}]{sutjin}
Suto, Y., Jing, Y.P., 1997, ApJS, 110, 167
\bibitem[\protect\citeauthoryear{Strauss,Ostriker \& Cen}{1998}]{striker}
Strauss, M.A.,Ostriker, J.P.,Renyue, C.,1998,\apj,494,20
\bibitem[\protect\citeauthoryear{Somerville,Primack \& Nolthenius}{1997}]{somr}
{{Somerville}, R.~S. and {Primack}, J. and {Nolthenius}, R.},\apj,1997,479,606
\bibitem[\protect\citeauthoryear{Somerville,Davis \& Primack}{1997}]{somdav}
Somerville, R.S.,Davis, M., \& Primack, J.R.,1997,\apj,479,616
\bibitem[\protect\citeauthoryear{Taylor \& Hamilton}{1996}]{tayl}
Taylor, A. N. \& Hamilton, A. J. S. 1996, MNRAS, 282, 767
\bibitem[\protect\citeauthoryear{Tadros \& Efstathiou}{1996}]{tadefsta}
Tadros, H. and Efstathiou, G.,1996,MNRAS,282,1381
\bibitem[\protect\citeauthoryear{Verde et al.}{2002}]{verd}
Verde, L.~et al.\  ,2002, \mnras, 335, 432 
\bibitem[\protect\citeauthoryear{White}{2001}]{white}
White, M.,2001,\mnras,321,1 
\bibitem[\protect\citeauthoryear{Willick et al.}{1997}]{uilik}
Willick, J.A., Strauss, M.A., Dekel, A., \& Kollat, T.,1997,\apj,486,629
\bibitem[\protect\citeauthoryear{Zehavi et al.}{2002}]{jevi}
Zehavi, I. et al.,2002,\apj,571,172
\bibitem[\protect\citeauthoryear{Zurek et al.}{1994}]{jur}
Zurek, W.H.,Quinn, P.J.,Salmon, J.K., \& Warren, M.S.,1994,\apj,431,559
\end{thebibliography}
\end{document}